\documentclass{IEEEcsmag}

\usepackage[colorlinks,urlcolor=blue,linkcolor=blue,citecolor=blue]{hyperref}

\usepackage{upmath}
\usepackage{amssymb,amsfonts}
\usepackage{amsmath}
\usepackage{graphicx}
\usepackage{multirow,multicol}
\usepackage{subfigure}
\usepackage{textcomp}
\usepackage{booktabs,tabularx,tabu}
\usepackage{color}
\usepackage{threeparttable}
\usepackage[superscript]{cite}
\usepackage{stackengine}
\usepackage{lipsum}

\newcommand\blfootnote[1]{%
  \begingroup
  \renewcommand\thefootnote{}\footnote{#1}%
  \addtocounter{footnote}{-1}%
  \endgroup
}

\newsavebox\mybox

\makeatletter
\newcommand*{\rom}[1]{\expandafter\@slowromancap\romannumeral #1@}

\makeatother

\makeatletter
\DeclareRobustCommand{\textsupsub}[2]{{%
  \m@th\ensuremath{%
    ^{\mbox{\fontsize\sf@size\z@#1}}%
    _{\mbox{\fontsize\sf@size\z@#2}}%
  }%
}}
\makeatother

\jvol{XX}
\jnum{XX}
\paper{8}
\begin{document}

\title{DHNet: Double MPEG-4 Compression Detection via Multiple DCT Histograms}

\author{Seung-Hun Nam \hfill Jihyeon Kang}
\affil{NAVER WEBTOON AI \hspace{44mm} NAVER WEBTOON AI}
\author{Wonhyuk Ahn \hspace{55mm} In-Jae Yu}
\affil{NAVER WEBTOON AI \hspace{44mm} Samsung Electronics}
\author{Myung-Joon Kwon}
\affil{Korea Advanced Institute of Science and Technology}

\begin{abstract}
In this article, we aim to detect the double compression of MPEG-4, a universal video codec that is built into surveillance systems and shooting devices.
Double compression is accompanied by various types of video manipulation, and its traces can be exploited to determine whether a video is a forgery.
To this end, we present a neural network-based approach with discriminant features for capturing peculiar artifacts in the discrete cosine transform (DCT) domain caused by double MPEG-4 compression.
By analyzing the intra-coding process of MPEG-4, which performs block-DCT-based quantization, we exploit multiple DCT histograms as features to focus on the statistical properties of DCT coefficients on multiresolution blocks.
Furthermore, we improve detection performance using a vectorized feature of the quantization table on dense layers as auxiliary information.
Compared with neural network-based approaches suitable for exploring subtle manipulations, the experimental results reveal that this work achieves high performance.

\begin{keywords}
Video forensics, Double MPEG-4 compression detection, Multiple DCT histograms, Auxiliary feature, Compression artifacts
\end{keywords}
\vspace{-8mm}
\end{abstract}

\maketitle

\chapterinitial{Multimedia forensics} is a field of research aimed at verifying the integrity of multimedia content and detecting fine-grained artifacts caused by multimedia forgery.\blfootnote{\copyright~2022 IEEE. Personal use of this material is permitted. Permission from IEEE must be obtained for all other uses, in any current or future media, including reprinting/republishing this material for advertising or promotional purposes, creating new collective works, for resale or redistribution to servers or lists, or reuse of any copyrighted component of this work in other works.} 
Due to the rapid proliferation of shooting devices and user-friendly editing software (e.g., Adobe Photoshop and Premiere Pro), the volume of forged content is increasing, and multimedia forensic research is becoming ever more significant.
Usually, manipulations are performed so exquisitely that their traces are difficult to distinguish via the human visual system (HVS); thus, forensic techniques have been designed to explore and capture subtle changes.
A fundamental assumption underlying multimedia forensics is that intrinsic statistical fingerprints of pristine content, such as acquisition and compression artifacts, are different from forged ones {\cite{verdoliva2020media}}.
Noting this point of view, forensic researchers have developed techniques on various media, including image, video, and audio.

\begin{figure}[t]
\centering{\includegraphics[width=1.0\linewidth]{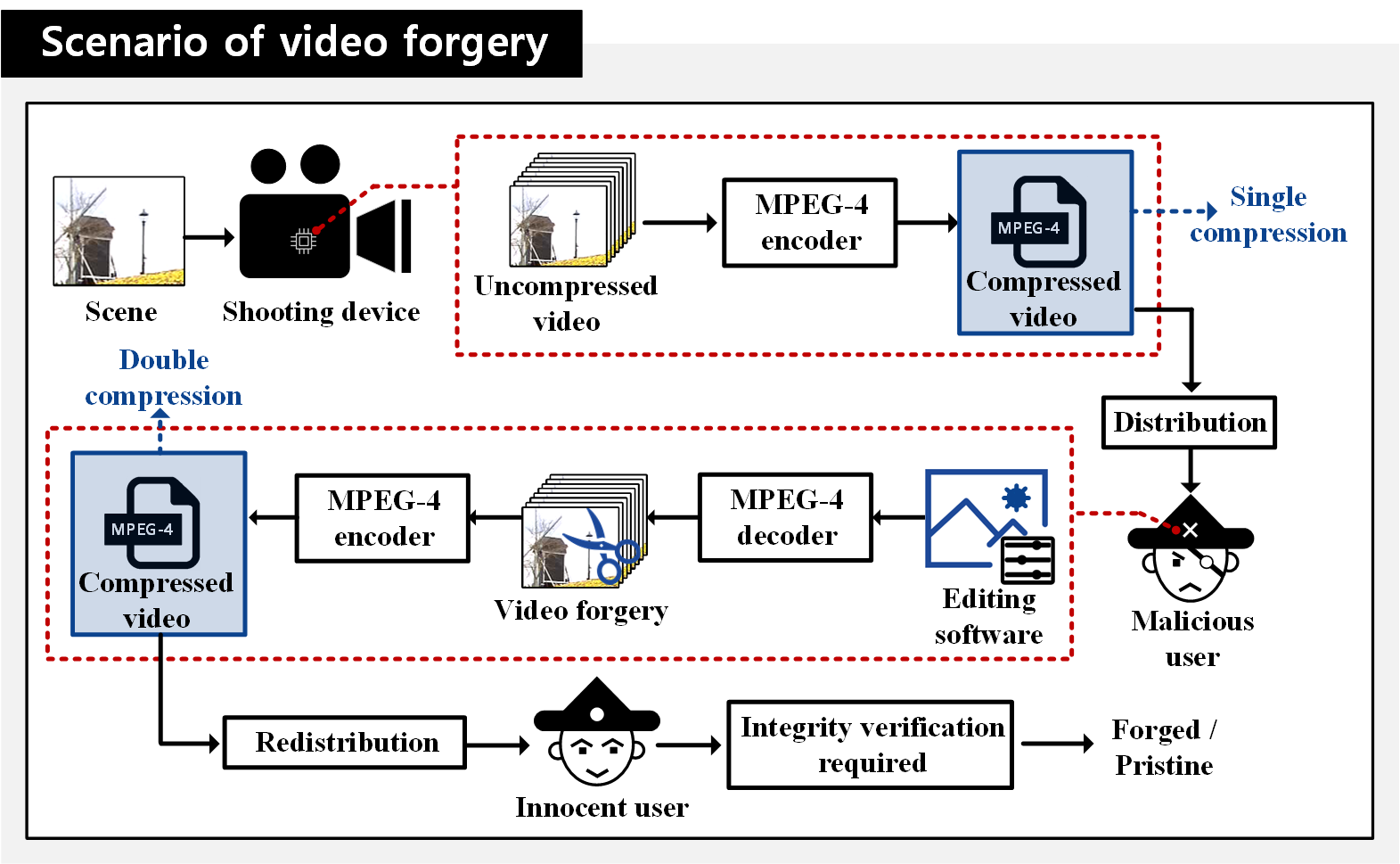}}
\caption{Universal scenario of double compression accompanying video forgery.}

\label{fig_relation_forgery_DC}
\end{figure}

Recently, under the influence of the advances in video-sharing platforms, research on video forensics has been conducted, with the following various targets: double compression, frame-rate conversion, and inter-frame forgery.
As illustrated in Figure~\ref{fig_relation_forgery_DC}, double compression is a weighty indicator of video forgery {\cite{henet}}.
Due to the storage issue, uncompressed videos are generally single-compressed before distribution, and when videos have been tampered with by an editing tool on the malicious distributor side, a second compression occurs during the storage process.
As the video forgery process commonly involves double compression, detecting recompression traces helps verify the integrity of a video {\cite{hvgg}}.

Focusing on this issue, many researchers aimed to detect double compression in video and chose MPEG-4 part 2 (hereinafter referred to as MPEG-4) as the target codec, a typical lossy compression standard developed by the MPEG group and widely adopted in surveillance systems, shooting devices, and editing tools.
MPEG-4 is largely composed of intra and predictive coding, which are used to reduce spatial and temporal redundancy, respectively {\cite{mp4_neuro}}.
In particular, intra coding is applied to the intra-coded frame (I-frame) located in the first of each group of pictures (GOP), where quantization is performed on the discrete cosine transform (DCT) domain of each divided $8\times8$ block.
When considering double compression, if the first and second quantization tables have differences, the distribution of the corresponding DCT coefficients differs from that of the single-compressed media {\cite{dcth_barni}}.

In the past, approaches with handcrafted features (e.g., Markov statistics-based features {\cite{mp4_markov}}) were dominant to distinguish the minute statistical variations on the decoded I-frame.
After the superior performance of the data-driven approach using a convolutional neural network (CNN) was demonstrated, network architecture, allocating preprocessing layer {\cite{henet}} or unpooled layers {\cite{srnet}} in the initial layers to capture low-level signals, was proposed.
Recently, approaches using both the CNN and handcrafted features (e.g., DCT histogram {\cite{dcth_barni}} and vectorized quantization matrix {\cite{dcth_park}}) demonstrated excellent performance in detecting double compression in JPEG files.
Since intra coding is based on a JPEG-like scheme, it is crucial to investigate the efficacy of these features in terms of detecting double MPEG-4 compression.

Motivated by the mentioned issue, neural network-based video forensics (referred to as DHNet) using discriminant features is presented.
To explore and learn distinguishable footprints caused by block-DCT-based double quantization, we devise multiple DCT histograms and auxiliary features suitable for capturing subtle statistical characteristics of the DCT coefficients.
Extensive experiments with large datasets demonstrated that our method outperformed comparable methods.
The main contributions are as follows:
\begin{itemize}
    \item To the best of our knowledge, this is the first attempt to employ multiple DCT histograms, which are designed to focus on the statistical properties of the DCT coefficients of multiresolution blocks as features for detecting double MPEG-4 compression.
    \item By analyzing the differences between MPEG-4 and JPEG in terms of quantization table acquisition, we propose auxiliary features with the base quantization table and quantizer scale, which contribute to the performance improvement of DHNet.
\end{itemize}

The remainder of this paper is organized as follows:
In “Background” section, we review the intra coding of MPEG-4, cases of double MPEG-4 compression, and the previous works related to this work.
Then, “Proposed Method” section introduces the proposed forensic approach utilizing distinctive features, followed by the performance analysis in “Experiments” section.
Finally, the article is concluded in “Conclusion” section.

\begin{figure*}[t]
\centering{\includegraphics[width=0.96\linewidth]{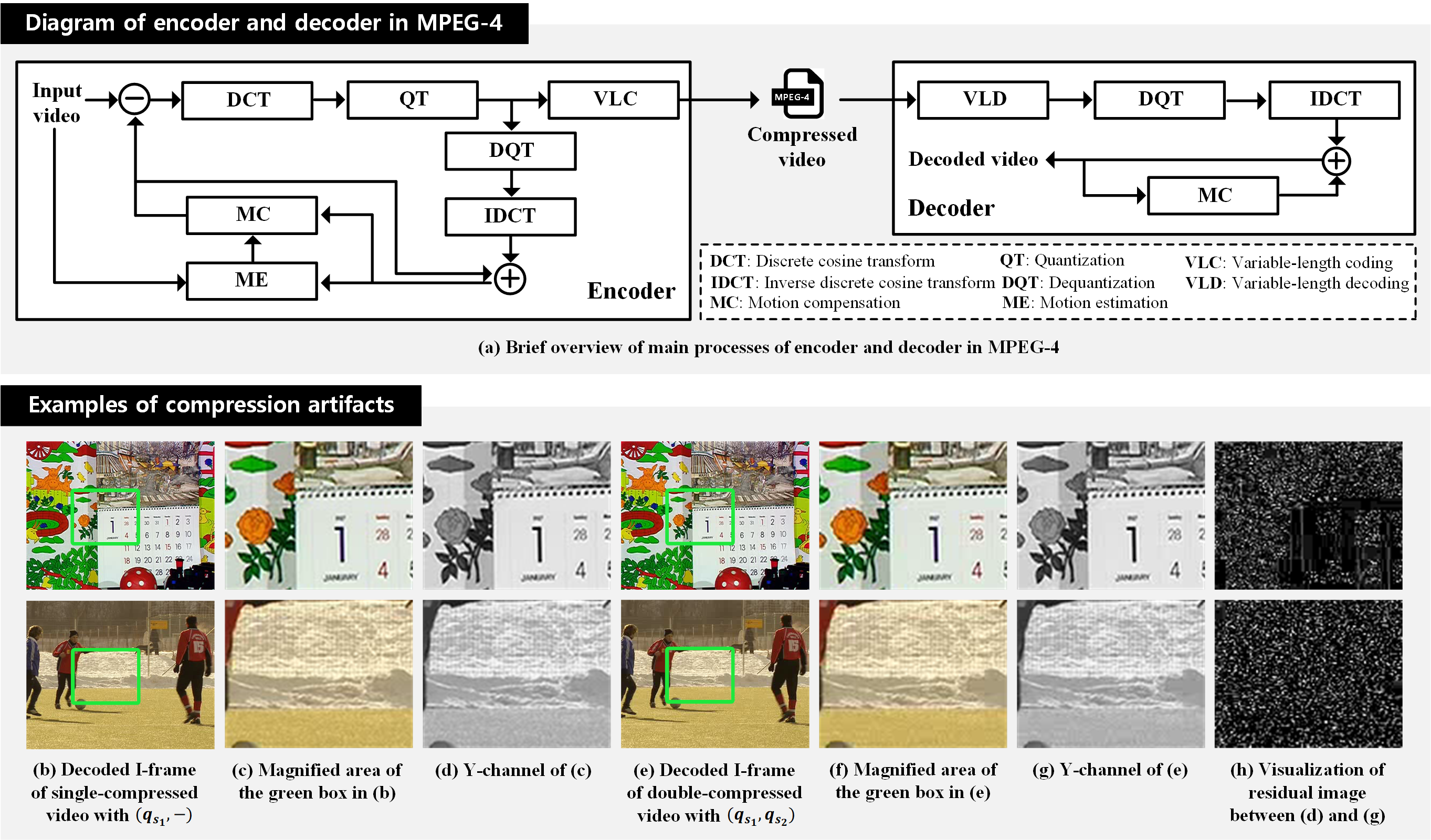}}
\caption{Block diagram of MPEG-4 and examples of the visualization of the residual between decoded single- and double-compressed I-frames, where $q_{s_1}=3$ and $q_{s_2}=5$.}
\label{fig_mpeg4}
\vspace{-4mm}
\end{figure*}

\section{BACKGROUND}

\subsection{Analysis of Intra Coding in MPEG-4}

For the MPEG-4 standard, the GOP is exploited as the base encoding unit, and each GOP comprises an I-frame, predictive frame (P-frame), and bidirectionally predictive frame (B-frame).
The I-frames are coded using intra coding based on DCT and quantization to reduce spatial redundancy (see Figure~\ref{fig_mpeg4}(a)).
The P-frames are predicted from previous I- or P-frames exploiting motion estimation and compensation to reduce temporal redundancy.
In this work, B-frames encoded from past and future frames are not considered for simplicity, and compression quality is controlled in variable bitrate (VBR) mode for efficient bitrate allocation.
In VBR mode, quality is determined by the quantizer scale $q_{s}$, and a value closer to 1 indicates the highest level.

Examples of compression artifacts remaining in decoded I-frame, caused by double compression, are illustrated in Figures~\ref{fig_mpeg4}(b)–\ref{fig_mpeg4}(h).
First, Figures~\ref{fig_mpeg4}(b) and \ref{fig_mpeg4}(e) show examples of decoded I-frames of single- and double-compressed videos with the configuration of ($q_{s_1},-$) and ($q_{s_1},q_{s_2}$), respectively.
No significant difference was observed by the HVS; however, when observing partially textured and flat areas of enlarged images (Figures~\ref{fig_mpeg4}(c) and \ref{fig_mpeg4}(f)), subtle differences can be identified.
To focus on these fine distortions, we visualized the residual image between the y-channel of the enlarged images and observed perceptually distinct artifacts (see Figure~\ref{fig_mpeg4}(h)).
These footprints, distinguishable from single compression, occur in rounding errors that are inevitably caused in the double quantization of intra coding.

\begin{figure*}[t]
\centering{\includegraphics[width=0.96\linewidth]{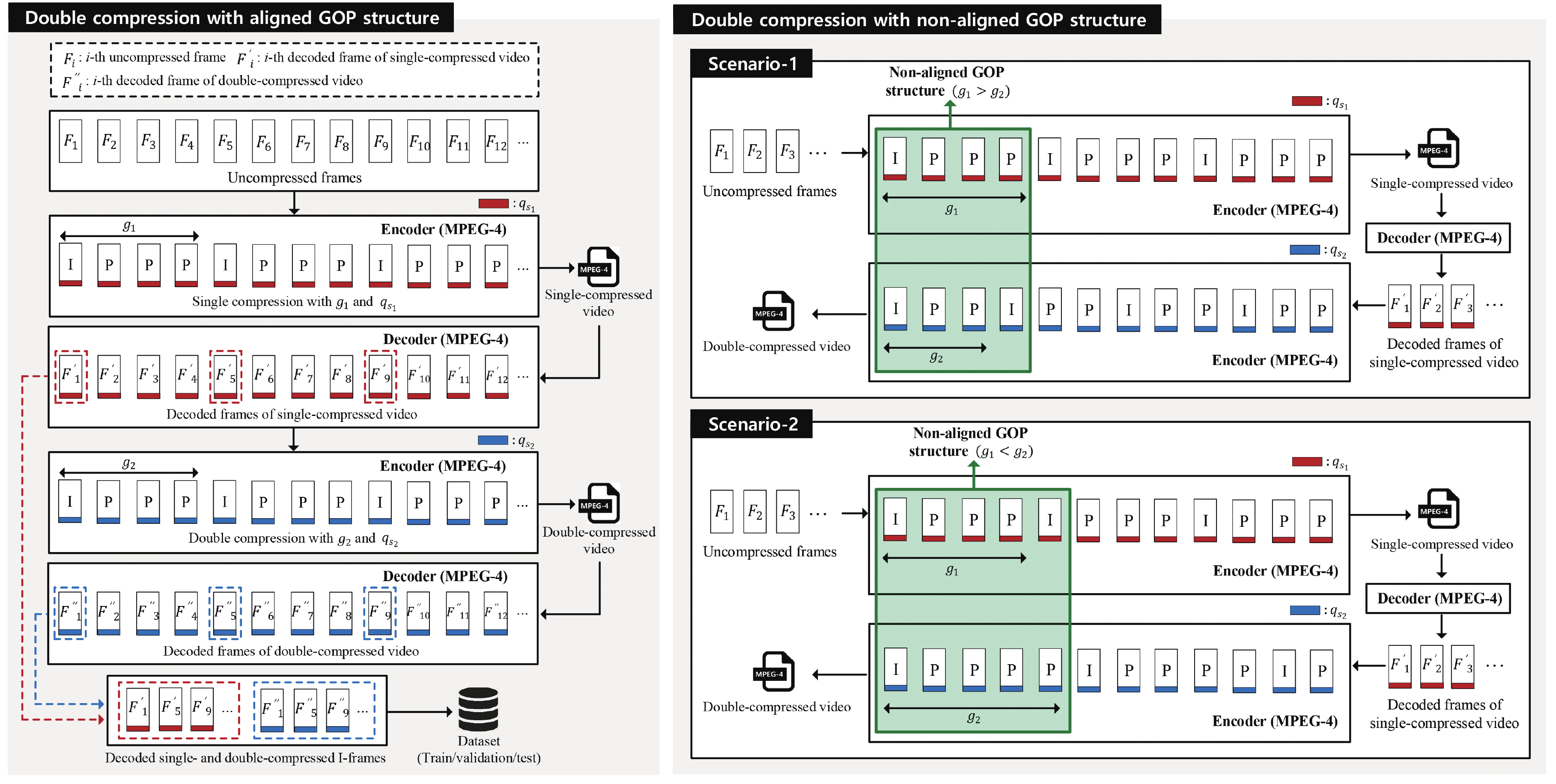}}
\vspace{-2mm}
\caption{Illustrative examples of double MPEG-4 compression with aligned and non-aligned GOP structures.}
\label{fig_scenario}
\vspace{-3mm}
\end{figure*}

In detail, intra coding divides a frame into $8\times 8$ blocks and then quantizes the DCT coefficients of each block with $q_{s}$ and the default quantization matrix $Q_{m}\in \mathbb{N}^{8\times8}$.
We let $C(i,j)$ and $C_{q}(i,j)$ be block-DCT coefficients and quantized coefficients at coordinates $(i,j)$, respectively.
As depicted in work{\cite{mp4_markov}}, the quantization process is as follows:
\vspace{-1mm}
\[
\resizebox{\linewidth}{!}{$
C_{q}(i,j)=(C(i,j)\times S(Q_{m}(i,j), q_{s})+r)\gg(s_{b}-3), (1)
$}
\vspace{-1mm}
\]
where $S(\cdot)$, $r$, $\gg x$, and $s_{b}$ denote the function to generate a scaled quantization matrix, the rounding value, a right-shift by $x$ bits on binary numbers, and the default scale-bits, respectively.
The dequantized coefficients $C_{d}(i,j)$ can be generated from the dequantization process as follows
\vspace{-2mm}
\[
\resizebox{1.0\linewidth}{!}{$
C_{d}(i,j)=(C_{q}(i,j)\times Q_m(i,j)\times q_{s})\gg 3. \quad\quad\quad (2)
$}
\vspace{-1mm}
\]
Unlike JPEG, the elements of the fixed quantization table in intra coding are prescaled due to $q_s$; hence, the divisor relationship between $q_{s_1}$ and $q_{s_2}$ is an important issue for tracing the rounding error caused by double quantization{\cite{mp4_markov}}.
This article excludes the case in which $q_{s_1}$ = $q_{s_2}$ in consideration of this constraint, which is a more realistic condition.
By analyzing intra coding, we identified the following important points.

\color{black}

\begin{itemize}

\item First, as intra coding employs JPEG-like approaches, we were inspired to adopt a DCT histogram suitable for detecting double JPEG.
In addition, multiple histograms of multiresolution blocks may help solve a given task in that they comprehensively explore the statistical properties of the DCT domain. 
\item Secondly, $q_{s}$ and $Q_{m}$ are essential parameters in intra coding; hence, they can be usefully exploited as auxiliary features to help a CNN learn double compression artifacts.
\vspace{-3mm}
\end{itemize}

\color{black}

\color{black}
\subsection{Case Study of Double Compression}

In VBR mode, $q_{s}$ and GOP size $g$ are key configurations of MPEG-4.
As presented in Figure \ref{fig_scenario}, the following double compression cases occur according to their combination: aligned and non-aligned GOP structures.
First, for the aligned GOP structure ($g_{1}=g_{2}$), quantization-based intra coding is periodically applied to the first frame of each GOP; hence, double quantization with a different quantization scale inevitably causes rounding errors, leaving distinguishable artifacts against single compression{\cite{mp4_markov}} (see the dashed red and blue boxes in Figure \ref{fig_scenario}).
Next, for the non-aligned GOP structure ($g_{1}\neq g_{2}$), inferring the index of the double-compressed I-frame in a given video except for the first frame is a challenging task.
To handle this issue, we trained a model with datasets of massive decoded I-frames generated for the aligned case and then aimed to infer the first I-frame of the non-aligned case to the pretrained model.
The countermeasures for non-aligned cases are detailed covered in the experimental section.

\begin{figure*}[t]
\centering{\includegraphics[width=0.96\linewidth]{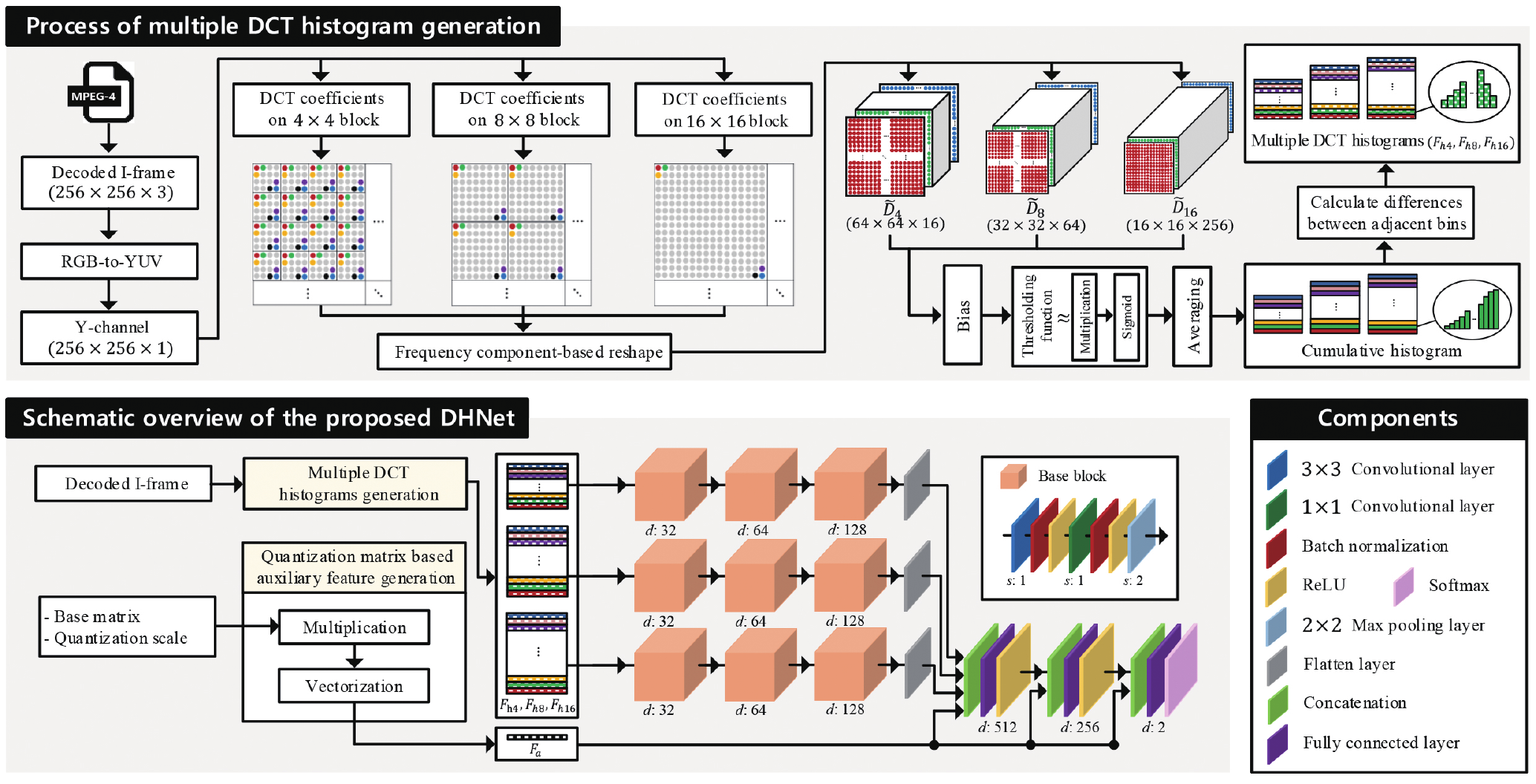}}
\vspace{-2mm}
\caption{Schematic overview of the proposed DHNet with multiple DCT histograms and auxiliary features: $d$ and $s$ indicate the number of output feature maps and stride of each layer, respectively.}
\vspace{-2mm}
\label{fig_schematic_overview}
\vspace{-1mm}
\end{figure*}

\color{black}
\vspace{-1mm}
\subsection{Related Work}

Conventional approaches for detecting double compression were based on handcrafted features and statistical characterization to explore fingerprints caused by the complicated processes of the target codec {\cite{verdoliva2020media}}.
To capture fine artifacts, Sun \emph{et al.}{\cite{sun2012exposing}} and He \emph{et al.}{\cite{mp4_neuro}} employed the first digit distribution of quantized coefficients and block artifacts combined with the variation of prediction footprints, respectively.
Jiang \emph{et al.} presented a linear discriminant analysis-based method using the Markov feature in the DCT domain {\cite{mp4_markov}}.
Aghamaleki and Behrad employed feature vectors obtained from the GOP and support vector machines {\cite{mp4_qm}}.
Yao \emph{et al.} used temporal segmentation to capture abnormal artifacts in rapid and static periods {\cite{yao2020double}}.
Currently, video forensics targeting high-efficiency codec (e.g., HEVC) is being actively studied.
Moreover, approaches based on the distribution of the DCT domain in a transform unit {\cite{li2019detection}}, analysis of the quality degradation mechanism in intra coding {\cite{jiang2019detection}}, and analysis of the quality degradation process in inter coding{\cite{xu2021detection}} were presented.

With the advances in CNNs, Boroumand \emph{et al.}{\cite{srnet}}, Nam \emph{et al.}{\cite{lfnet}}, and Yoon \emph{et al.}{\cite{yoon2021frame}} demonstrated that placing an unpooled layer on the initial layer helps capture low-level signals (e.g., stego signals, local artifacts, and interpolation artifacts).
With the help of the preprocessing with a high-pass filter, He \emph{et al.}{\cite{henet}} and Nam \emph{et al.}{\cite{hvgg}} proposed CNNs to classify relocated I-frames and double-compressed I-frames, respectively.
To detect double JPEG compression accompanying an $8\times8$ unit quantization, such as intra coding in MPEG-4, Barni \emph{et al.} used a CNN-based approach exploiting the DCT histogram of all DC and AC components {\cite{dcth_barni}}.
Park \emph{et al.} improved upon the method {\cite{dcth_barni}} by providing a quantization table in the header to the CNN, which exhibited performance improvement {\cite{dcth_park}}.
Kwon \emph{et al.} presented a JPEG artifact learning module (J-ALM) using the DCT volume representation and quantization matrix to learn the compression artifacts {\cite{kwon2021cat}}.

Inspired by the previous works, we devise discriminant features to capture peculiar artifacts and design multi-stream architecture that effectively learns those features.

\section{PROPOSED METHOD}
Figure \ref{fig_schematic_overview} presents the whole procedure of the proposed approach consisting of the following three steps: multiscale DCT histogram generation, auxiliary feature generation, and multistream CNN-based feature learning.

\subsection{Preliminaries}
Let video \textbf{v} with resolution $W$$\times$$H$ be represented by \textbf{v} $=$ \{$...,f_{t},...$\}, where $f_{t}$ denotes a \emph{t}-th decompressed frame in RGB format.
The proposed approach uses the decompressed I-frame indicated by $I$ $=$ \{$f_{t_{I}}$\} $\in$ $\mathbb{Z}^{W\times H\times3}$ as input.
Here, $t_{I}$ denotes the index of the I-frame.
The preprocessed input is defined as $I_{y}$ $=$ \{$Y(I)$\} $\in$ $\mathbb{R}^{W\times H\times1}$, where $Y(\cdot)$ denotes the Y-channel of the input after RGB-to-YUV conversion.

\subsection{Multiple DCT Histogram Generation}
Inspired by an approach{\cite{dcth_barni}} using a single histogram to detect double JPEG, we devised multiple DCT histograms as features to focus on the statistical properties of DCT coefficients on multiresolution blocks of sizes $4\times4$, $8\times8$, and $16\times16$.
For simplicity, the size of each block is defined as $\delta\times\delta$, where $\delta=\{4,8,16\}$, and the generated DCT histogram for each block size is denoted as $F_{h_{\delta}}$.
This subsection focuses on the $F_{h_{\delta}}$ generation process, and the detailed procedure is illustrated in Figure \ref{fig_schematic_overview}.
First, the DCT coefficients of each $\delta\times \delta$ block of $I_{y}$ are computed.
For frequency-specific analysis, the obtained coefficients are reshaped into $\tilde{D}_\delta$ (size of $W/\delta \times H/\delta \times \delta^{2}$) with the same frequency component for each channel.
$\tilde{D}_{c}$, which indicates each channel of $\tilde{D}_\delta$, is the output of a 2D convolution operation with stride $\delta$ between $I_{y}$ and $H_{c}$, where $H_{c}$ is a 2D DCT basis for frequency $c$, $c\in\{(1,1),(1,2),...,(\delta,\delta)\}$.

For each channel, a cumulative histogram is calculated based on $\tilde{D}_{c}$ and the boundary value $b$ of the bin in the histogram, where $b\in\{-\alpha,-\alpha+1,...,\alpha\}$.
As depicted in work{\cite{dcth_barni}}, $B_{c,b}$ is the $b$-th bin in a cumulative histogram for $c$ and it means the average number of values in $\tilde{D}_{c}$ that are greater than $b$.
$B_{c,b}$ is computed as follows:
\[
\resizebox{\linewidth}{!}{$
B_{c,b}=\frac{\delta^{2}}{W\times H}\sum_{j=1}^{H/\delta}\sum_{i=1}^{W/\delta}{T(\tilde{D}_{c}(i,j)-b)}, \quad\quad (3)
$}
\]
where $T(\cdot)$ indicates the threshold function that turns positive and negative numbers into 1 and 0, respectively.
Computing $B_{c,b}$ for all $c$ and $b$ completes the cumulative histogram and DCT histogram feature $F_{h}$ can be generated by calculating the difference between adjacent bins in the cumulative histogram: $F_{h_\delta} =\{f|f_{c,b}=B_{c,b+1}-B_{c,b}, \forall c, b\}$, where the size of $F_{h_\delta}$ is $2\alpha \times \delta^{2} \times 1$.
Figure \ref{fig_schematic_overview} reveals an intuitive explanation of histogram generation, and we used multiple DCT histograms generated by following the mentioned procedures.

\begin{figure*}[t]
\centering{\includegraphics[width=0.96\linewidth]{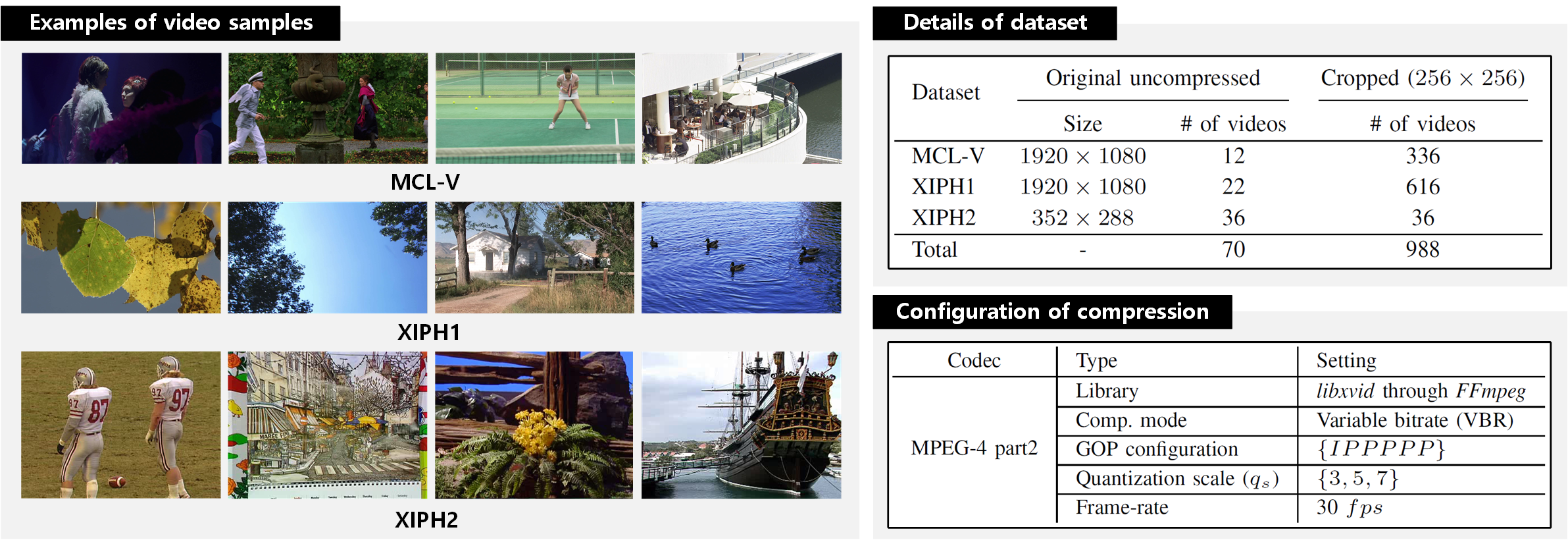}}
\vspace{-1mm}
\caption{Examples of the content of video samples constituting each dataset, such as MCL-V{\cite{mclv}}, XIPH1{\cite{xiph}}, and XIPH2{\cite{xiph}}, and details of employed videos dataset and MPEG-4 compression configuration.}
\label{fig_video_examples}
\vspace{-2mm}
\end{figure*}

\vspace{-2mm}

\color{black}
\subsection{Auxiliary Feature Generation}
The variables $Q_{m}$ and $q_{s}$ indicate a quantization matrix of $8\times8$ and the quantizer scale value of the last compression, respectively.
Unlike JPEG, in which a quantization table corresponding to compression quality is stored in the header, quantization and dequantization on intra coding are conducted based on $Q_{m}$ and $q_{s}$.
Inspired by previous work{\cite{dcth_park}}, we expected that these include useful information when the CNN learns double compression artifacts; hence, we devised the auxiliary feature $F_{a}$ based on $Q_{m}$ and $q_{s}$.
The proposed feature $F_{a}$ can be computed as: $F_{a}=V(Q_{m})\times q_{s}$, where $V(\cdot)$ denotes a vectorization function and the size of $F_{a}$ is $64\times 1$.
In this work, we considered and investigated the following two types of tables:
\vspace{-1mm}
\[
\resizebox{\linewidth}{!}{$
  Q_{m_1}=
  \left[\begin{smallmatrix}
    1 & 1 & 1 & 1 & 1 & 1 & 1 & 1 \\
    1 & 1 & 1 & 1 & 1 & 1 & 1 & 1 \\
    1 & 1 & 1 & 1 & 1 & 1 & 1 & 1 \\
    1 & 1 & 1 & 1 & 1 & 1 & 1 & 1 \\
    1 & 1 & 1 & 1 & 1 & 1 & 1 & 1 \\
    1 & 1 & 1 & 1 & 1 & 1 & 1 & 1 \\
    1 & 1 & 1 & 1 & 1 & 1 & 1 & 1 \\
    1 & 1 & 1 & 1 & 1 & 1 & 1 & 1 \\
  \end{smallmatrix}\right], 
  Q_{m_2}= 
  \left[\begin{smallmatrix}
    8&16&19&22&26&27&29&34 \\
    16&16&22&24&27&29&34&37 \\
    19&22&26&27&29&34&34&38 \\
    22&22&26&27&29&34&37&40 \\
    22&26&27&29&32&35&40&48 \\
    26&27&29&32&35&40&48&58 \\
    26&27&29&34&38&46&56&69 \\
    27&29&35&38&46&56&69&83 \\
  \end{smallmatrix}\right], (4)$} 
\]
where $Q_{m_1}$ is a matrix of ones which has no contextual meaning, and $Q_{m_2}$ is a default quantization table on intra coding.
In particular, $Q_{m_1}$ and $Q_{m_2}$ were expected to help the CNN in $q_{s}$-focused learning and quantization process inference, respectively.
In this article, the CNNs employing $Q_{m_1}$ and $Q_{m_2}$ as auxiliary features are referred to as CNN-Q1 and CNN-Q2, respectively.

\subsection{Network Architecture}

We designed a multistream CNN using multiple DCT histograms of $F_{h_4}$, $F_{h_8}$, $F_{h_{16}}$ and the auxiliary feature $F_{a}$ as input.
As illustrated in Figure \ref{fig_schematic_overview}, the generated DCT histograms are input into each stream, consisting of three base blocks.
Each base block consists of $3\times3$ or $1\times1$ convolutional (Conv) layers, each of which is followed by batch normalization, and a rectified linear unit (ReLU).
The $3\times3$ Conv layer learns the relationship between neighboring elements, and the $1\times1$ Conv layer learns the association between sequentially placed feature maps.
The last layer of each block was applied with $2\times2$ max pooling with a stride of 2 to reduce the dimensionality of the feature maps.
After flattening the outputs of the last base block of each stream, the flattened feature was concatenated and input into the fully connected (FC) layers block, consisting of the FC layers, ReLU, and softmax.
To induce DHNet to learn auxiliary features, we concatenated the generated $F_{a}$ with flattened features and the activations of two FC layers.
The output of the last FC layer was input into a two-way softmax to predict the class label.

\vspace{-1mm}

\subsection{Loss Function}
We define the loss function as follows: $L=L_{c}+L_{r}$, where $L_{c}$ and $L_{r}$ are the cross-entropy loss and regularization term, respectively.
$L_{c}$ is computed as: $L_{c} = -(1-l)\times log(\frac{e^{y_{0}}}{e^{y_{0}}+e^{y_{1}}}) - l\times log(\frac{e^{y_{1}}}{e^{y_{0}}+e^{y_{1}}})$, where the output of the network is $y=[y_{0};y_{1}]$ ($l= 0$ if the $I_{y}$ is single-compressed and $l= 1$ if the $I_{y}$ is double-compressed).
To reduce the chance of overfitting, $L_{r}$ is defined as: $L_{r}=\gamma\sum_{n=1}^{N}{{||w_{n}||}^{2}_{2}}$, where $w_{n}$ represents the weight matrix of the $n$-th Conv layer.

\begin{table*}[t]
\centering
\caption{Performance comparison with comparative methods for classifying double MPEG-4 compression.}
\begin{threeparttable}
\centering
\begin{subtable}{}
\vspace{-2mm}
{(a) Experimental results on frame-wise detection.}

\resizebox{\linewidth}{!}{%\scriptsize
\begin{tabular}{l c c c c c c c c c c c c c c c}
\specialrule{1.0pt}{1.0pt}{1.0pt}
\multirow{2}{*}{Method} & \multirow{2}{*}{ACC $\uparrow$} & \multirow{2}{*}{TNR $\uparrow$} & \multirow{2}{*}{PRE $\uparrow$} &\multirow{2}{*}{REC $\uparrow$} &\multirow{2}{*}{F1-S $\uparrow$} & \multirow{2}{*}{AUC $\uparrow$} & \multicolumn{3}{c}{ACC on single comp. $(q_{s_1},-)$} & \multicolumn{6}{c}{ACC on double comp. $(q_{s_1},q_{s_2})$}\\
\cmidrule(lr){8-10} \cmidrule(lr){11-16}
& & & & & & & $(3,-)$ & $(5,-)$ & $(7,-)$ & $(3,5)$ & $(3,7)$ & $(5,3)$ & $(5,7)$ & $(7,3)$ & $(7,5)$ \\
\hline 
HeNet \cite{henet} & 67.81 & 54.47 & 64.06 & 81.15 & 71.60 & 0.7372 & 96.51 & 52.01 & 14.89 & 73.82 & 90.72 & 49.34 & 91.42 & 87.80 & 93.82 \\
H-VGG \cite{hvgg} & 69.90 & 51.60 & 64.57 & 88.20 & 74.56 & 0.7852 & 97.72 & 45.31 & 11.77 & 75.03 & 89.05 & 81.25 & 93.71 & 92.68 & 97.48 \\
FCDNet \cite{yoon2021frame} & 80.07 & 74.35 & 76.98 & 85.79 & 81.15 & 0.8951 & 99.77 & 83.25 & 40.03 & 68.57 & 85.03 & 79.54 & 93.6 & 90.63 & 97.37   \\
SRNet \cite{srnet} & 80.08 & 74.60 & 77.11 & 85.56 & 81.12 & 0.8805 & 97.31 & 76.80 & 49.71 & 76.01 & 84.54 & 76.60 & 90.62 & 89.48 & 96.11 \\
LFNet \cite{lfnet} & 81.48 & 76.65 & 78.71 & 86.31 & 82.33 & 0.9051 & 97.51 & 80.02 & 52.42 & 74.62 & 80.08 & 83.22 & 92.80 & 89.71 & 97.48 \\
BarniNet \cite{dcth_barni} & 85.78 & 84.03 & 84.57 & 87.53 & 86.03 & 0.9469 & 97.65 & 81.88 & 72.57 & 84.68 & 82.74 & 83.54 & 90.97 & 89.48 & 93.82 \\
ParkNet \cite{dcth_park}  & 86.32 & 87.62 & 87.29 & 85.02 & 86.14 & 0.9450 & 98.11 & 85.42 & 79.33 & 81.14 & 76.46 & 83.65 & 88.19 & 88.00 & 92.68\\
ParkNet-Q1 \cite{dcth_park} & 90.21 & 91.22 & 91.04 & 89.21 & 90.12 & 0.9569 & 99.31 & 91.08 & 83.27 & 78.51 & 77.71 & 98.51 & 86.28 & 99.31 & 94.97 \\
ParkNet-Q2 \cite{dcth_park}  & 90.15 & 91.81 & 91.53 & 88.49 & 89.98 & 0.9618 & 97.71 & 88.07 & 89.65 & 79.20 & 73.25 & 98.85 & 84.11 & 99.54 & 96.03 \\
J-ALM-Q1 \cite{kwon2021cat} & 91.68 & 92.43 & 92.32 & 90.93 & 91.62 & 0.9718 & 99.66 & 94.34 & 83.31 & 76.69 & 85.14 & 93.71 & 94.06 & 99.20 & 96.80\\
J-ALM-Q2 \cite{kwon2021cat} & 91.23 & 92.28 & 92.11 & 90.17 & 91.13 & 0.9672 & 99.23 & 94.21 & 83.40 & 75.66 & 83.77 & 93.37 & 92.23 & 99.54 & 96.46 \\
DHNet & 88.51 & 87.78 & 87.95 & 89.23 & 88.59 & 0.9553 & 98.97 & 86.74 & 77.63 & 85.02 & 85.60 & 84.43 & 92.68 & 91.20 & 96.45 \\
DHNet-Q1 & 92.53 & 93.32 & 93.22 & 91.74 & 92.47 & 0.9768 & 99.48 & 92.05 & 88.45 & 80.34 & 82.05 & 98.51 & 91.31 & 99.77 & 98.51 \\
DHNet-Q2 & 92.37 & 92.98 & 92.90 & 91.76 & 92.33 & 0.9587 & 99.54 & 91.31 & 88.11 & 84.57 & 80.88 & 96.57 & 90.97 & 99.77 & 97.82 \\
\specialrule{1.0pt}{1.0pt}{1.0pt}
\end{tabular}
}
\end{subtable}
\end{threeparttable}

\vspace{3mm}

\begin{threeparttable}
\centering
\begin{subtable}{}
\vspace{-2mm}
{(b) Experimental results on GOP-wise detection.}
\resizebox{\linewidth}{!}{%\scriptsize
\centering
\begin{tabular}{c l c c c c c c c c c c c c c c c}
\specialrule{1.0pt}{1.0pt}{1.0pt}
\multirow{2}{*}{$\Phi$} & \multirow{2}{*}{Method} & \multirow{2}{*}{ACC $\uparrow$} & \multirow{2}{*}{TNR $\uparrow$} & \multirow{2}{*}{PRE $\uparrow$} &\multirow{2}{*}{REC $\uparrow$} &\multirow{2}{*}{F1-S $\uparrow$} & \multirow{2}{*}{AUC $\uparrow$} & \multicolumn{3}{c}{ACC on single comp. $(q_{s_1},-)$} & \multicolumn{6}{c}{ACC on double comp. $(q_{s_1},q_{s_2})$}\\
\cmidrule(lr){9-11} \cmidrule(lr){12-17}
& & & & & & & & $(3,-)$ & $(5,-)$ & $(7,-)$ & $(3,5)$ & $(3,7)$ & $(5,3)$ & $(5,7)$ & $(7,3)$ & $(7,5)$ \\
\hline 

\multirow{12}{*}{5} & FCDNet \cite{yoon2021frame} & 81.84 & 71.42 & 76.35 & 92.26 & 83.56 & 0.9211 & 100 & 89.14 & 25.14 & 79.52 & 96.25 & 82.29 & 98.82 & 96.65 & 100 \\
& SRNet \cite{srnet} & 82.89 & 75.76 & 78.78 & 90.02 & 84.03 & 0.9105 & 98.59 & 80.55 & 48.14 & 87.36 & 88.67 & 79.07 & 92.70 & 94.51 & 97.78\\
& LFNet \cite{lfnet} & 85.66 & 78.85 & 81.39 & 92.46 & 86.57 & 0.9342 & 98.85 & 85.71 & 52.00 & 85.06 & 82.93 & 89.13 & 98.85 & 98.84 & 100 \\
& BarniNet \cite{dcth_barni} & 88.71 & 85.90 & 86.65 & 91.53 & 89.03 & 0.9650 & 99.43 & 84.57 & 73.71 & 85.90 & 91.57 & 88.78 & 94.38 & 93.67 & 94.89 \\
& ParkNet \cite{dcth_park}  & 91.04 & 90.67 & 90.74 & 91.41 & 91.07 & 0.9758 & 100 & 90.29 & 81.71 & 90.80 & 85.39 & 87.78 & 91.36 & 94.25 & 98.90 \\
& ParkNet-Q1 \cite{dcth_park}  & 93.39 & 95.24 & 95.05 & 91.55 & 93.27 & 0.9911 & 100 & 96.00 & 89.71 & 80.17 & 79.01 & 100 & 92.22 & 100 & 97.87\\
& ParkNet-Q2 \cite{dcth_park}  & 93.19 & 95.42 & 95.22 & 90.96 & 93.04 & 0.9828 & 98.86 & 94.29 & 93.14 & 86.81 & 75.28 & 98.79 & 86.05 & 100 & 98.87 \\
& J-ALM-Q1 \cite{kwon2021cat} & 94.31 & 93.70 & 93.78 & 94.92 & 94.35 & 0.9906 & 100 & 96.54 & 84.57 & 84.00 & 91.40 & 96.43 & 97.72 & 100 & 100  \\
& J-ALM-Q2 \cite{kwon2021cat} & 93.74 & 93.95 & 93.93 & 93.52 & 93.72 & 0.9874 & 100 & 96.57 & 85.29 & 83.53 & 90.72 & 95.00 & 91.89 & 100 & 100  \\
& DHNet & 92.76 & 91.43 & 91.64 & 94.10 & 92.85 & 0.9857 & 100 & 94.86 & 79.43 & 89.87 & 93.26 & 88.64 & 96.39 & 95.51 & 100 \\
& DHNet-Q1 & 95.52 & 96.00 & 95.97 & 95.05 & 95.51 & 0.9945 & 100 & 96.57 & 91.43 & 86.02 & 89.89 & 100 & 95.23 & 100 & 100 \\
& DHNet-Q2 & 94.57 & 95.05 & 95.00 & 94.10 & 94.55 & 0.9892 & 100 & 93.71 & 91.43 & 84.52 & 89.13 & 97.59 & 94.05 & 100 & 98.90 \\
\hline

\multirow{12}{*}{10} & FCDNet \cite{yoon2021frame} & 83.60 & 74.86 & 78.60 & 92.34 & 84.92 & 0.9379 & 100 & 94.86 & 29.71 & 78.47 & 93.07 & 87.37 & 98.83 & 96.30 & 100  \\
& SRNet \cite{srnet} & 84.37 & 76.95 & 79.93 & 91.79 & 85.45 & 0.9195 & 99.01 & 82.33 & 49.51 & 90.00 & 88.61 & 81.25 & 94.12 & 97.87 & 98.90\\
& LFNet \cite{lfnet} & 86.81 & 81.14 & 83.06 & 92.47 & 87.51 & 0.9485 & 100 & 88.71 & 54.71 & 83.13 & 82.95 & 90.70 & 98.01 & 100 & 100 \\
& BarniNet \cite{dcth_barni} & 90.11 & 88.76 & 89.06 & 91.45 & 90.24 & 0.9760 & 100 & 88.02 & 78.27 & 91.56 & 85.23 & 84.42 & 96.04 & 92.71 & 98.75 \\
& ParkNet \cite{dcth_park}  & 92.48 & 92.57 & 92.56 & 92.39 & 92.47 & 0.9804 & 100 & 91.43 & 86.29 & 89.89 & 85.11 & 91.21 & 88.10 & 100 & 100\\
& ParkNet-Q1 \cite{dcth_park}  & 95.74 & 97.90 & 97.81 & 93.58 & 95.65 & 0.9889 & 100 & 98.86 & 94.86 & 86.14 & 85.19& 100 & 91.25& 98.94& 100\\
& ParkNet-Q2 \cite{dcth_park}  & 93.82 & 96.95 & 96.75 & 90.68 & 93.61 & 0.9869 & 100 & 93.14 & 97.71 & 88.10 & 77.63 & 97.09 & 85.91 & 97.47 & 97.88\\
& J-ALM-Q1 \cite{kwon2021cat} & 95.76 & 95.80 & 95.79 & 95.71 & 95.75 & 0.9926 & 100 & 97.10 & 90.29 & 86.53 & 93.18 & 95.70 & 1000 & 98.86 & 100  \\
& J-ALM-Q2 \cite{kwon2021cat} & 94.71 & 95.88 & 95.78 & 93.54 & 94.65 & 0.9924 & 100 & 96.50 & 91.14 & 80.04 & 91.11 & 95.45 & 96.94 & 100 & 97.70   \\
& DHNet & 94.49 & 94.67 & 94.65 & 94.32 & 94.48 & 0.9913 & 100 & 95.43 & 88.57 & 90.00 & 93.76 & 87.64 & 97.80 & 96.70 & 100\\
& DHNet-Q1 & 97.12 & 97.90 & 97.87 & 96.34 & 97.10 & 0.9950 & 100 & 96.00 & 97.71 & 88.31 & 93.40 & 98.87 & 97.47 & 100 & 100\\
& DHNet-Q2 & 96.31 & 97.14 & 97.10 & 95.49 & 96.29 & 0.9941 & 100 & 95.43 & 96.00 & 92.13 & 89.47 & 96.39 & 96.10 & 100 & 98.88 \\

\specialrule{1.0pt}{1.0pt}{1.0pt}
\end{tabular}
}
\end{subtable}
\end{threeparttable}
\vspace{-2mm}
\label{table_framewise}
\end{table*}

\section{EXPERIMENTS}
\subsection{Experimental Settings}
\subsubsection{Datasets}
We employed uncompressed videos from the following datasets: MCL-V~{\cite{mclv}}, XIPH1~{\cite{xiph}}, and XIPH2~{\cite{xiph}}.
Examples of video samples constituting each dataset, information on the details of the videos, and configuration of compression are depicted in Figure~\ref{fig_video_examples}.
Before the first compression, we obtained 988 cropped videos with the resolution $256 \times256$ using a non-overlapped crop in the spatial domain with various content.
The videos were divided into three sets for training, validation, and testing (with an $8:1:1$ ratio).
For compression, \textit{libxvid} in \textit{FFmpeg} was used, and the GOP structure was set to \{IPPPPP\}.
$q_{s_1}$ and $q_{s_2}$, which control the quality of the first and second compressions, were set to $\{3,5,7\}$ in VBR mode.
Based on the literature {\cite{mp4_markov}}, it was not considered double compression when $q_{s_1}=q_{s_2}$.

With these parameters, 2,964 ($988\times3$) single-compressed videos and 5,928 ($988\times3\times2$) double-compressed videos were obtained.
To train the DHNet and evaluate the performance of frame-wise detection, we obtained 120,000 decoded I-frames from the generated videos where the ratio of positive to negative samples is 1:1.
The $q_s$ and base quantization tables for auxiliary features were stored in the database along with the generated decoded frames.
In the experiments on temporal and GOP-wise detection, the single- and double-compressed videos generated for testing were used.
The setting for response to non-aligned GOP structure (i.e., $g_{1}\neq g_{2}$) is introduced in the last subsection.

\subsubsection{Baselines}

We compared the proposed DHNet with the following eight CNN-based methods for multimedia forensics and steganalysis: HeNet {\cite{henet}}, H-VGG~{\cite{hvgg}}, FCDNet~{\cite{yoon2021frame}}, SRNet~{\cite{srnet}}, LFNet~{\cite{lfnet}}, BarniNet~{\cite{dcth_barni}}, ParkNet~{\cite{dcth_park}}, and J-ALM~{\cite{kwon2021cat}}.
For HeNet and H-VGG, high-pass filtering (e.g., $\frac{1}{12}[-1, 2, -2, 2, -1; 2, -6, 8, -6, 2;-2,8,-12,$ $8,-2; 2,-6,8,-6,2;-1,2,-2,2,-1]$) was used as the preprocessing layer to focus on the traces remaining in a frame.
For BarniNet and ParkNet, a histogram based on an $8\times8$ block-DCT was used as input.
The auxiliary features were provided as optional for ParkNet, but were provided as essential for J-ALM.
Furthermore, we employed MF-LDA~{\cite{mp4_markov}}, a non-CNN approach using five GOPs for the performance comparison of the GOP-wise detection.

\begin{figure*}[t]
\centering{\includegraphics[width=0.96\linewidth]{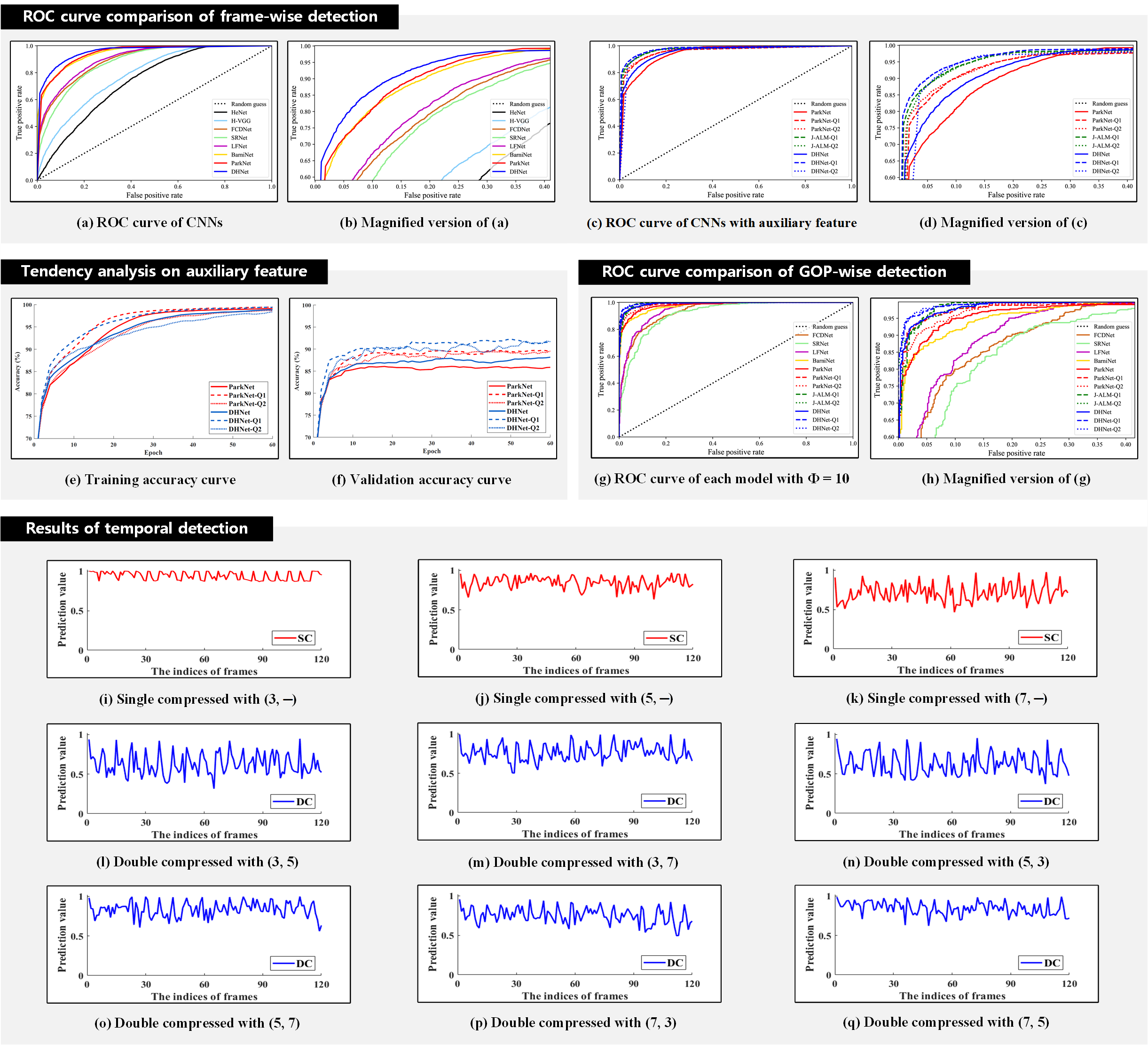}}
\caption{Comparison of the receiver operating characteristic (ROC) curve, accuracy tendencies of training and validation phases on auxiliary features, and temporal detection results of DHNet-Q1 for each frame along the temporal axis.}
\vspace{-1mm}
\label{fig_graph}
\end{figure*}

\subsubsection{Evaluation Metrics}

Detection performance was measured using the accuracy (ACC), true negative rate (TNR), precision (PRE), recall (REC), and F1-score (F1-S).
Each metric is defined as follows: ACC (\%) $= \frac{TP+TN}{TP+TN+FP+FN}\times100$, TNR (\%) $= \frac{TN}{TN+FP}\times100$, PRE (\%) $= \frac{TP}{TP+FP}\times100$, REC (\%) $= \frac{TP}{TP+FN}\times100$, and F1-S (\%) $= \frac{2\times PRE \times REC}{PRE+REC}\times100$, where $TP$, $TN$, $FP$, and $FN$ are the numbers corresponding to the true positives, true negatives, false positives, and false negatives, respectively.
In addition, the receiver operating characteristic (ROC) curves were computed to compare the performance of the proposed and comparative methods.
The ROC curve is defined as a plot of the true positive rate against the false positive rate, and the area under the curve (AUC) is further used as a metric.

\subsubsection{Implementation Details}

Using the TensorFlow framework, we built the network and conducted experiments using NVIDIA GeForce RTX 2080 Ti.
We used the Adam optimizer with a learning rate of $10^{-4}$, momentum coefficients $\beta_{1}$ = 0.9 and $\beta_{2}$ = 0.999, and the numerical stability constant $\epsilon$ = $10^{-8}$.
The mini-batch size, $\alpha$, and $\gamma$ were set to 32, 60, and $10^{-4}$, respectively.
The selected model maximized the detection accuracy on the validation set for 60 epochs.

\vspace{-1mm}

\subsection{Results on Frame-wise Detection}
First, we conducted frame-wise detection with comparative CNN-based methods, which are specialized in exploring quantization artifacts on block-DCT or detecting fine-grained signals.
Table \ref{table_framewise}(a) lists the performance metrics for each model for frame-wise detection, and we computed the ROC curve for a detailed performance analysis (see Figures~\ref{fig_graph}(a) and \ref{fig_graph}(b)).
Without the help of auxiliary features, the ACC, TNR, PRE, REC, F1-S, and AUC of the DHNet were 88.51\%, 87.78\%, 87.95\%, 89.23\%, 88.59\%, and 0.9553, respectively.
Importantly, DCT histogram-based approaches, including BarniNet~{\cite{dcth_barni}}, ParkNet~{\cite{dcth_park}}, and our DHNet, effectively detected double compression artifacts on the decoded I-frame, achieving an ACC of over 85\%.
Particularly, among them, DHNet exhibited the best performance for all metrics, suggesting that the proposed method of adopting multiple DCT histograms better explores the statistical properties remaining in the DCT domain than comparative CNNs.

Next, to test the leverage of auxiliary features, we observed the performance change in the presence or absence of $F_a$.
Providing the auxiliary features with the base matrix $Q_m$ and the quantizer scale $q_s$ to the dense layers enhanced the detection performance of DHNet and ParkNet (see Figures~\ref{fig_graph}(c) and \ref{fig_graph}(d)).
Importantly, the comprehensively analyzed results of Table \ref{table_framewise}(a) reveal that the DHNet-Q1 performs better than competitive methods.
In addition, J-ALM-Q1 specialized in capturing the compression artifacts showed second-best performance.
As the detection performance of CNN-Q1 outperforms that of CNN-Q2 overall, it is interpreted that $q_s$ is more helpful in identifying double compression than the elements constituting $Q_{m_2}$.
Furthermore, accuracy tendency curves for training and validation in Figures~\ref{fig_graph}(e) and \ref{fig_graph}(f) indicate that the adopting auxiliary features helps the model converge efficiently.
These curves confirmed that the devised additional features facilitate single- and double-compression assessment.

\begin{figure*}[t!]
\centering{\includegraphics[width=0.96\linewidth]{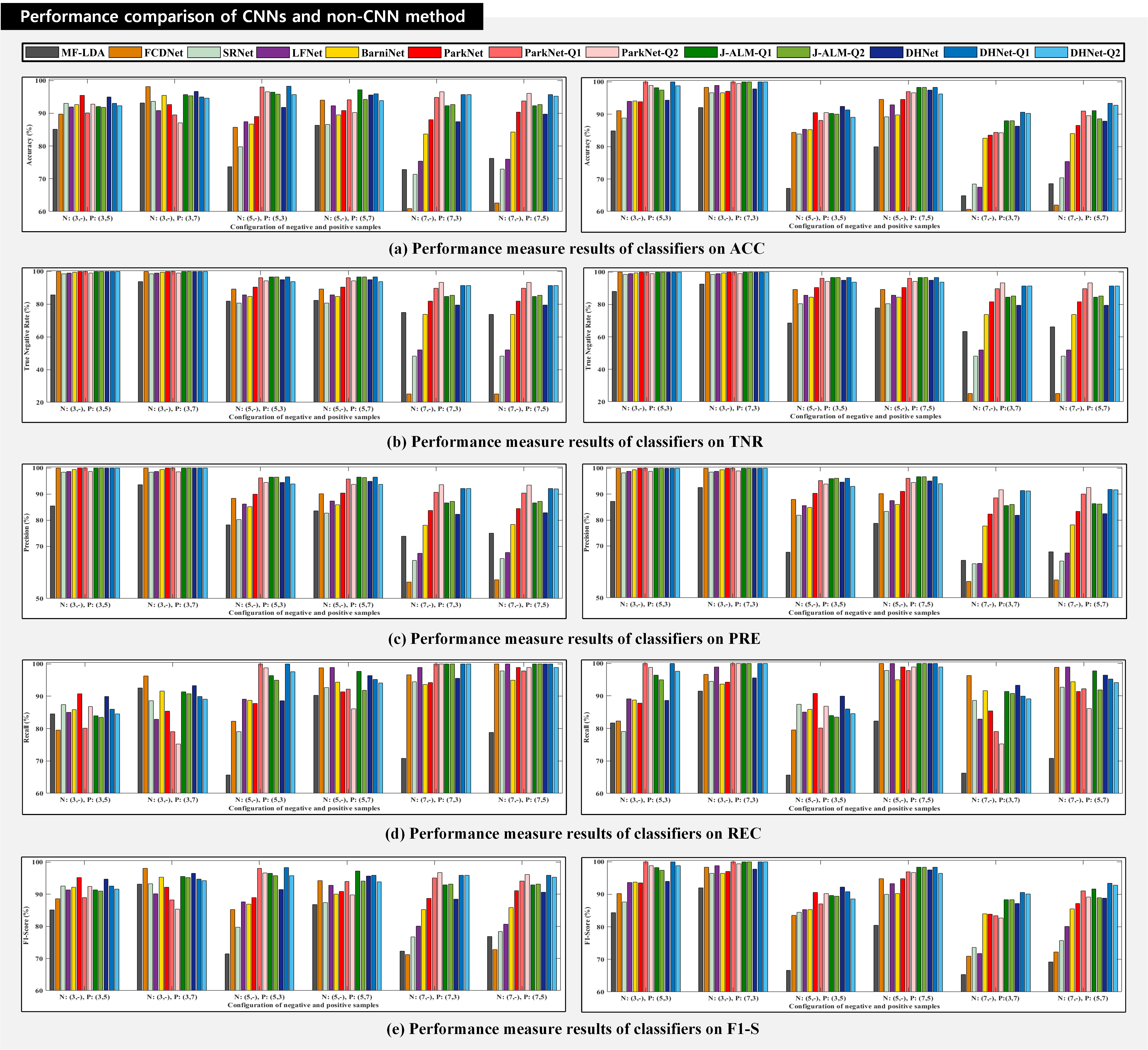}}
\vspace{-1mm}
\caption{Performance comparison with CNNs and non-CNN method on single- and double-compressed pairs.}
\vspace{-2mm}
\label{fig_result_MF-LDA}
\end{figure*}

\vspace{-1mm}

\subsection{Results on GOP-wise Detection}

As mentioned in the subsection on datasets, videos for the main experiment were generated with a fixed GOP configuration; hence, peculiar quantization traces were repeatedly found in the decoded I-frames of the double-compressed video (Figure \ref{fig_scenario}).
To verify this experimentally using DHNet-Q1, we performed temporal detection on each frame constituting a sample video (from the first to the 120\textsuperscript{th} frame).
As displayed in Figures~\ref{fig_graph}(i) to \ref{fig_graph}(q), on both single- and double-compressed videos, the prediction value peaks were generated periodically in GOP size intervals.
In particular, the detection performance relatively deteriorated in inference to the decoded P-frames.

\vspace{-1mm}

Next, we conducted GOP-wise detection on the test videos generated with the aligned GOP structure.
Noting the temporal detection results according to the frame index, only decoded I-frames extracted from $\Phi$ GOPs on a given video were provided to the pretrained model.
In this experiment, the top seven models for frame-wise detection were employed, and the majority voting mechanism was applied to the $\Phi$ prediction results.
When $\Phi$ was set to 5 and 10, the detection ACC values of the DHNet-Q1 were 95.52\% and 97.14\%, respectively, as presented in Table \ref{table_framewise}(b).
As the number of I-frames provided to each model increased (i.e., $\Phi=1$ $\rightarrow$ $\Phi=5,10$), the number of correct predictions for the corresponding true class increased.

\vspace{-1mm}

Furthermore, we performed a performance comparison with the non-CNN approach, MF-LDA~{\cite{mp4_markov}} (see Figure \ref{fig_result_MF-LDA}).
Unlike the CNN-based approaches to testing with only a single trained model, MF-LDA requires classifiers for each parameter and a combination of a single compression and double compression pair (requiring 12 classifiers for MF-LDA). The Markov features used by MF-LDA exhibited degraded performance in discriminating the statistical differences between specific pairs, and our work demonstrated outstanding performance overall.

\subsection{Additional Analysis}
\subsubsection{Ablation Study}
To verify the effectiveness of multiple DCT histograms, we performed an ablation study on the combination of the histograms of $F_{h_4}$, $F_{h_8}$, and $F_{h_{16}}$.
Seven combinations of DCT histograms were used, and the stream composition of the DHNet was adjusted according to the number of input features. When three types of DCT histograms were input into the DHNet, it displayed the best accuracy of 88.51\%. For other combinations, accuracy values of less than 88\% were observed. As the number of histograms provided to the DHNet decreased, the detection performance tended to deteriorate.

\vspace{-2mm}

\subsubsection{Results on the Non-aligned GOP Structures}

As illustrated in Figure \ref{fig_scenario}, a non-aligned GOP structure may occur on double compression in videos.
We experimented to determine the double compression of a given video using only the first decoded I-frame to cover this issue.
The pretrained DHNet-Q1 without retraining was exploited, and videos were generated for the following GOP configurations $(g_1,g_2)$: $(5,6)$, $(6,5)$, $(6,7)$, and $(7,6)$.
The average accuracy value was measured at 92.45\%, which is 0.08\% lower performance than the results for the aligned GOP structure (see Table \ref{table_framewise}(a)).
The traces of the double quantization considered in the model training phase are left in the first I-frame of the video; thus, it is possible to handle the assumed case of a non-aligned structure.

\vspace{-1mm}

\section{CONCLUSION}

In this article, we proposed a novel DHNet that identifies double MPEG-4 compression.
For MPEG-4, block-DCT-based double compression with a different quantization scale inevitably causes rounding errors, leaving distinguishable footprints against single compression.
Therefore, we devised multiple DCT histograms suitable for capturing peculiar artifacts in the DCT domain.
Furthermore, performance improvement was achieved by providing auxiliary features to dense layers.
The results revealed that this work is superior to the competitive methods, including CNN and non-CNN methods, in terms of various objective metrics.
In the future, we will investigate the efficacy of the devised discriminant features against various compression standards that utilize block-based quantization.

\section{ACKNOWLEDGMENT}
This research was developed with supporting from NAVER WEBTOON AI.

\bibliographystyle{IEEEtran}

\begin{thebibliography}{10}

\bibitem{verdoliva2020media}
L.~Verdoliva, ``Media forensics and deepfakes: an overview,'' \emph{IEEE J.
  Sel. Top. Signal Process.}, vol.~14, no.~5, pp. 910--932, 2020.

\bibitem{henet}
P.~He \emph{et~al.}, ``Frame-wise detection of relocated i-frames in double
  compressed h. 264 videos based on convolutional neural network,'' \emph{J.
  Vis. Commun. Image Represent.}, vol.~48, pp. 149--158, 2017.

\bibitem{hvgg}
S.-H. Nam, J.~Park, D.~Kim, I.-J. Yu, T.-Y. Kim, and H.-K. Lee, ``Two-stream
  network for detecting double compression of h. 264 videos,'' in \emph{proc.
  IEEE Int. Conf. Image Process.}, 2019, pp. 111--115.

\bibitem{mp4_neuro}
P.~He, X.~Jiang, T.~Sun, and S.~Wang, ``Detection of double compression in
  mpeg-4 videos based on block artifact measurement,'' \emph{Neurocomputing},
  vol. 228, pp. 84--96, 2017.

\bibitem{dcth_barni}
M.~Barni \emph{et~al.}, ``Aligned and non-aligned double jpeg detection using
  convolutional neural networks,'' \emph{J. Vis. Commun. Image Represent.},
  vol.~49, pp. 153--163, 2017.

\bibitem{mp4_markov}
X.~Jiang, W.~Wang, T.~Sun, Y.~Q. Shi, and S.~Wang, ``Detection of double
  compression in mpeg-4 videos based on markov statistics,'' \emph{IEEE Sign.
  Process. Letters}, vol.~20, no.~5, pp. 447--450, 2013.

\bibitem{srnet}
M.~Boroumand, M.~Chen, and J.~Fridrich, ``Deep residual network for
  steganalysis of digital images,'' \emph{IEEE Trans. Info. Forens. Security},
  2018.

\bibitem{dcth_park}
J.~Park, D.~Cho, W.~Ahn, and H.-K. Lee, ``Double jpeg detection in mixed jpeg
  quality factors using deep convolutional neural network,'' in \emph{Proc.
  Eur. Conf. Comput. Vis.}, 2018, pp. 636--652.

\bibitem{sun2012exposing}
T.~Sun, W.~Wang, and X.~Jiang, ``Exposing video forgeries by detecting mpeg
  double compression,'' in \emph{proc. IEEE Int. Conf. Acoustics, Speech and
  Signal Processing}, 2012, pp. 1389--1392.

\bibitem{mp4_qm}
J.~A. Aghamaleki and A.~Behrad, ``Detecting double compressed mpeg videos with
  the same quantization matrix and synchronized group of pictures structure,''
  \emph{J. Electron. Imaging}, vol.~27, no.~1, p. 013031, 2018.

\bibitem{yao2020double}
H.~Yao, R.~Ni, and Y.~Zhao, ``Double compression detection for h. 264 videos
  with adaptive gop structure,'' \emph{Multimedia Tools Appl.}, vol.~79, no.~9,
  pp. 5789--5806, 2020.

\bibitem{li2019detection}
Q.~Li, R.~Wang, and D.~Xu, ``Detection of double compression in hevc videos
  based on tu size and quantised dct coefficients,'' \emph{IET Inf. Secur.},
  vol.~13, no.~1, pp. 1--6, 2019.

\bibitem{jiang2019detection}
X.~Jiang, Q.~Xu, T.~Sun, B.~Li, and P.~He, ``Detection of hevc double
  compression with the same coding parameters based on analysis of intra coding
  quality degradation process,'' \emph{IEEE Trans. Info. Forens. Security},
  vol.~15, pp. 250--263, 2019.

\bibitem{xu2021detection}
Q.~Xu, X.~Jiang, T.~Sun, and A.~C. Kot, ``Detection of hevc double compression
  with non-aligned gop structures via inter-frame quality degradation
  analysis,'' \emph{Neurocomputing}, vol. 452, pp. 99--113, 2021.

\bibitem{lfnet}
S.-H. Nam \emph{et~al.}, ``Deep convolutional neural network for identifying
  seam-carving forgery,'' \emph{IEEE Trans. Circuits Syst. Video Technol.},
  vol.~31, no.~8, pp. 3308--3326, 2021.

\bibitem{yoon2021frame}
M.~Yoon \emph{et~al.}, ``Frame-rate up-conversion detection based on
  convolutional neural network for learning spatiotemporal features,''
  \emph{arXiv:2103.13674}, 2021.

\bibitem{kwon2021cat}
M.-J. Kwon, I.-J. Yu, S.-H. Nam, and H.-K. Lee, ``Cat-net: Compression artifact
  tracing network for detection and localization of image splicing,'' in
  \emph{proc. IEEE/CVF Winter Conf. Appl. Comput. Vis.}, 2021, pp. 375--384.

\bibitem{mclv}
J.~Y. Lin \emph{et~al.}, ``Mcl-v: A streaming video quality assessment
  database,'' \emph{J. Vis. Commun. Image Represent.}, vol.~30, pp. 1--9, 2015.

\bibitem{xiph}
C.~Montgomery \emph{et~al.}, ``Xiph. org video test media (derf's collection),
  the xiph open source community, 1994,'' \emph{Online,
  https://media.xiph.org/video/derf}.

\end{thebibliography}

\vspace{2mm}

\begin{IEEEbiography}{Seung-Hun Nam}{\,} is currently working at NAVER WEBTOON AI, South Korea. He received the Ph.D. degree from the School of Computing, KAIST, South Korea, in 2020. His research interests include digital watermarking and multimedia forensics. Contact him at shnam1520@gmail.com.
\end{IEEEbiography}

\begin{IEEEbiography}{Wonhyuk Ahn}{\,} is currently working at NAVER WEBTOON AI, South Korea. He received the Ph.D. degree from the School of Computing, KAIST, South Korea, in 2021. His research interests include deep learning and multimedia content protection. Contact him at whahnize@gmail.com.
\end{IEEEbiography}

\begin{IEEEbiography}{Myung-Joon Kwon}{\,} is currently pursuing the Ph.D. degree from the School of Electrical Engineering, KAIST, South Korea. He received the M.S. degree from the School of Computing, KAIST, South Korea, in 2021. His research interests include adversarial attacks, computer vision, and machine learning. Contact him at mjkwon2021@gmail.com.
\end{IEEEbiography}

\begin{IEEEbiography}{Jihyeon Kang}{\,} is currently working at NAVER WEBTOON AI, South Korea. He received the Ph.D. degree from the Graduate School of Information Security, KAIST, South Korea, in 2021. His research interests include machine learning and computer vision. Contact him at mangorism@gmail.com.
\end{IEEEbiography}

\begin{IEEEbiography}{In-Jae Yu}{\,} is currently working at Samsung Electronics, South Korea. He received the Ph.D. degree from the School of Computing, KAIST, South Korea, in 2021. His research interests include machine learning and computer vision. Contact him at myhome98304@gmail.com.
\end{IEEEbiography}

\end{document}